\documentclass[onecolumn,showpacs,preprintnumbers,amsmath,amssymb]{revtex4}
\usepackage[english]{babel}
\usepackage{graphics}
\usepackage{epsf}
\usepackage{psfig}
\usepackage{dcolumn}
\usepackage{bm}
\begin{document}

\title{ Parametric Resonance in a Vibrating Cavity}

\author{Pawe\l{} W\c{e}grzyn}
\affiliation{ Marian Smoluchowski Institute of Physics,
Jagellonian University, Reymonta 4, 30-059 Cracow, Poland}
\email{wegrzyn@th.if.uj.edu.pl}

\begin{abstract}
We present the study of parametric resonance in a one-dimensional
cavity based on the analysis of classical optical paths. The
recursive formulas for field energy are given. We separate the
mechanism of particle production and the resonance amplification
of radiation. The production of photons is a purely quantum effect
described in terms of quantum anomalies in recursive formulas. The
resonance enhancement is a classical phenomenon of focusing and
amplifying beams of photons due to D\"{o}ppler effect.
\end{abstract}


\maketitle

\section{Introduction}

The phenomenon of parametric resonance is related to the
instability of open systems under the action of an external
periodic force. In quantum field theory,  we expect the resonant
amplification of quantum fluctuations \cite{berges,
hirooka,lambrecht}. The resonance enhancement of vacuum
fluctuations is usually referred as the dynamical Casimir effect.
The standard model to investigate this effect is the system of
electromagnetic fields confined inside a vibrating one-dimensional
cavity \cite{moore,fulling,dodonov,bordag}.

Meplan and Gignoux \cite{meplan} found the correspondence between
the wave propagation  in a vibrating cavity and the motion of
massless particles in a two-dimensional space-time billiard. They
used the generalized Korringa-Kohn-Rostoker method. The method
presented in this paper offers both a better insight in resonance
mechanism and a more useful tool for detailed calculations.
Moreover, the brief discussion of the quantum version in
\cite{meplan} does not trace properly  the impact of conformal
anomaly \cite{nagatani}.

In this Letter, we present a new approach to study resonance
solutions. We show that the resonance amplification together with
the formation of narrow packets in energy densities \cite{law} is
a purely classical phenomenon. It can be explained as a
consequence of a cumulative D\"{o}ppler effect \cite{dittrich}. In
this way, we are allowed to interpret a wave packet as a beam of
massless and non-interacting particles. Hence, we introduce the
billiard function which contains full information about possible
reflections from a cavity wall. This function is helpful to
establish recursive formulas for physical quantities in a
vibrating cavity. Production of particles in quantum case is
represented by conformal anomaly contribution in such formulas for
energy densities. Our technique is illustrated with solutions of
several examples of vibrating cavity systems discussed in
literature.

\section{Classical billiards}

Before studying the field-theoretic models, we consider briefly
the classical billiards.  Let us start with the simplest example,
namely a head-on collision of two non-relativistic objects. A
particle moves with velocities $v$ and $v'$ before and after the
collision, while  respective velocities of a target are denoted by
$u$ and $u'$. This elementary physical problem is fully described
by the reflection law in one dimension (on a line),
\begin{equation}
v+v'=u+u' \ .
\end{equation}
If we assume a target to be very heavy and skip its recoil, then
the velocity of a particle after the collision is just $v'=-v+2u$.
We see that if a particle were bounced with a regular frequency,
then its energy would grow quadratically with time. Let us look
now at a head-on collision of relativistic objects.  This two-body
problem is described by the relativistic reflection law which says
that the sums of rapidities  are equal,
\begin{equation}
{\rm artanh}\, v+{\rm artanh}\, v'={\rm artanh}\, u+{\rm artanh}\,
u' \ .
\end{equation}
In this case, if a particle bounces rhythmically on a heavy
target, then its rapidity grows linearly with time. It follows
that the energy grows exponentially with time. In particular, we
can consider  a left-moving massless particle (photon) hitting a
target with a velocity of light. The exact formula for the change
of the particle energy after a single collision yields:
\begin{equation}
E'=E\, \frac{1-u}{1+u}\left( 1+\frac{2E}{M}\sqrt{\frac{1-u}{1+u}}
\right)^{-1} \ .
\end{equation}
Two consecutive factors on the right-hand side are obviously the
D\"{o}ppler factor and the Compton factor. For a large target mass
$M$, the Compton factor is neglected. Therefore, during each
head-on collision the energy of a massless particle increases (or
decreases if the target moves in the same direction) by the
D\"{o}ppler factor.

\begin{figure}
\includegraphics{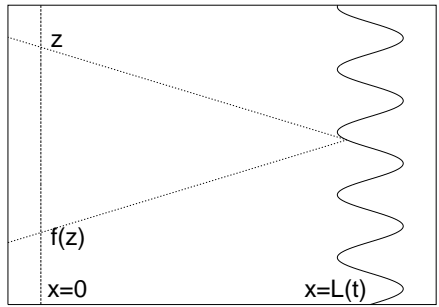}
\caption{The definition of the billiard function $f(z)$. It
returns full information about possible reflections  of a
light-like particle from the mirror moving with a prescribed
trajectory $L(t)$. The billiard function identifies the trajectory
uniquely and its derivative is the retarded D\"{o}ppler factor. }
\end{figure}

Now, we insert a heavy target (mirror) which moves with some
prescribed trajectory $L(t)$. Let us then introduce the following
{\it billiard function}\cite{wegrzyn}:
\begin{equation}\label{bf}
f(t+L(t))=t-L(t) \ .
\end{equation}

We define the retarded time $t^\star$ using the retardation
relation $t = t^\star+L(t^\star)$ and recognize that the
derivative of the billiard function is just the retarded
D\"{o}ppler factor:
\begin{equation}\label{df}
    \dot{f}(t)=\frac{1-\dot{L}(t^\star)}{1+\dot{L}(t^\star)} \ .
\end{equation}
Since the derivative is additive, the billiard function is
increasing and its inverse is well defined. The D\"{o}ppler factor
is greater than one (amplifying) if the mirror is moving left,
i.e. towards an arriving test massless particle. For any physical
trajectory of the mirror, its corresponding billiard function is
defined unambiguously. On the other hand, any proper billiard
function may be used to reconstruct its genuine mirror's
trajectory. More information can be found in \cite{wegrzyn}.
Although rather cumbersome in general, the billiard function is
the most useful characteristic to analyze the parametric resonance
within a particular cavity model. We can make use of the
correspondence between waves and massless particles and relate the
field theory problem with the analysis of trajectories in a
space-time billiard \cite{meplan}.

\section{Optical cavities}

Under the guidance of the above simple examples from classical
billiards, we consider the case of field theory. As usual, we will
discuss the ideal resonator model \cite{moore} formed by two
perfect mirrors. The electromagnetic potential $A(x,t)$  obeys the
one-dimensional wave equation and any classical solution is
composed of left- and right-moving wave packets,
\begin{equation}
A(t,x)=A_L(t+x)+A_R(t-x) \ .
\end{equation}
For the sake of simplicity, we assume the left mirror to be fixed
at $x=0$, whereas the right mirror is vibrating according to a
prescribed trajectory $L(t)$. Then, the field $A(x,t)$ is subject
to  Dirichlet's boundary conditions, namely:
\begin{equation}
A_L(\tau)= - A_R(\tau)\equiv \varphi(\tau)\  , \ \ \ \
\varphi(\tau)=\varphi(f(\tau)) \ . \label{dbc}
\end{equation}
We follow the solution of the Cauchy problem  with the help of the
billiard function. From initial data for $A$ and $\partial_t A$ at
$t=0$,  we read the function $\varphi$ for arguments in the
interval $[-L(0),L(0)]$ (up to some irrelevant constant). The
solution is obtained if we extend $\varphi$ into the whole real
domain. We can do it using Eq.~(\ref{dbc}) provided that the
billiard function is known. Usually, if we predetermine some
trajectory of the mirror, then the corresponding billiard function
is to be found only in the way of numerical iterations based on
Eq.~(\ref{bf}). On the other hand, to specify the mirror motion we
can  predetermine the billiard function instead.

 The energy density of the classical wave packet is given by:
\begin{equation}\label{ced}
    T_{00}(t,x)=1/2(\partial_t A)^2+1/2(\partial_x
    A)^2=\varrho(t+x)+\varrho(t-x) \ ,
\end{equation}
where $\varrho(\tau)=\dot{\varphi}^2(\tau)$. The formula for the
total energy can be  presented in the following form:
\begin{equation}\label{te}
E(t)=\int_0^{L(t)} \ dx \ T_{00}(t,x)=\int_{t-L(t)}^{t+L(t)} \
d\tau \ \varrho(\tau)  \ .
\end{equation}

The billiard function determines all possible trajectories of a
massless particle or all classical optical paths inside the
cavity. Each trajectory can be represented by the sequence
$\left\{ T_n(\tau)\right\}\equiv \left\{(f^{-1})^{\circ
n}(\tau)\right\}_{n=0}^{\infty}$, where $(f^{-1})^{\circ n}(\tau)$
denotes here $n$-fold composition $f^{-1}\circ f^{-1}\circ \ldots
\circ f^{-1}$. A light-like particle starts at time $\tau$ moving
right from $x=0$ and next elements of the sequence read next times
of collisions with
 the static mirror. We define also retarded times $T^\star_n(\tau)$ by demanding
$T^\star_n(\tau)+L(T^\star_n(\tau))=T_n(\tau)$. It is easy to find
the relation $T^\star_n(\tau)=(T_n(\tau)+T_{n-1}(\tau))/2$.

 The appearance of the parametric resonance is related to the
existence of periodic particle trajectories \cite{meplan}. Such
trajectories obey the following periodicity condition (for any
non-negative integer $n$):
\begin{equation}\label{pt}
     T_n(\tau_0)=\tau_0+n T \ ,
\end{equation}
where $\tau_0$ is a starting point in time  and $T$ is a period.
The retarded times are respectively
$T^\star_n(\tau_0)=T_n(\tau_0)-T/2$. Looking at  Fig.1, we notice
immediately that any periodic particle trajectory appears on
condition that the trajectory of the mirror has {\it return
points} \cite{cole}:
\begin{equation}\label{pc}
    L\left(\tau^\star_0+n T\right)=T/2 \ .
\end{equation}
It is usually assumed that the position of the mirror $T/2$ refers
to the length of the static (unperturbed) cavity, so that
$T/2\equiv L\equiv L(0)$ in this paper.

Let us define the characteristic for particle trajectories which
will play the most important role in our analysis. It is the
cumulative D\"{o}ppler factor:
\begin{equation}\label{cdf}
D_n(\tau)= \frac{1}{\dot{T}_n(\tau)} =\prod_{k=1}^n
\dot{f}(T_n(\tau))= \prod_{k=1}^n
\frac{1-\dot{L}(T_n^\star(\tau))}{1+\dot{L}(T_n^\star(\tau))} \ .
\end{equation}
We will call a trajectory to be {\it positive} (stable,
attractive) if its cumulative D\"{o}ppler factor tends to infinity
for large $n$, and to be {\it negative} (unstable, chaotic) if its
cumulative D\"{o}ppler factor goes to zero with increasing $n$.

The existence of periodic particle trajectories inside the cavity
is equivalent to the existence of suitable return points in the
trajectory of the cavity wall. In particular, a periodic particle
trajectory is positive (negative) if any return point of the
mirror trajectory gives  a D\"{o}ppler factor  greater (less) than
one. It happens if the right mirror is always moving toward
(outward) the left mirror at all return points.

Denote by $\tau_+$ and $\tau_-$ starting points for positive and
negative periodic trajectories respectively.
 For small perturbations $\varepsilon$,  it
is straightforward to derive the following formulas:
\begin{equation}\label{pert1}
T_n(\tau_++\varepsilon)\cong\tau_++n T+\varepsilon
D_n^{-1}(\tau_+) \ ,
\end{equation}
\begin{equation}\label{pert2}
T_n(\tau_-+\varepsilon D_n(\tau_-))\cong\tau_-+n T+\varepsilon
 \ .
\end{equation}

Now, we can show how the existence of  periodic particle
trajectories inside the cavity triggers the resonance instability
of the system. For the sake of simplicity, we assume that there
are only two periodic particle trajectories, a positive one and a
negative one, with starting points $\tau_+$ and $\tau_-$ lying in
the initial interval $[-L(0),L(0)]$. These simplifications are
only for clarity, and the method can be adapted for more
sophisticated examples. As it was said, to solve the Cauchy
problem we need to extend the function $\varphi$ given in
Eq.~(\ref{dbc}) from the initial interval to the whole real
domain. If we are interested only in the evolution forward in
time, then it is enough  to obtain data only for arguments to the
right of the initial interval $\tau>L(0)$.  To build such
extension of $\varphi$, we pick  points from the initial interval,
run along  particle trajectories and read  extended values of
$\varphi$ from Eq.~(\ref{dbc}). To analyze our solution obtained
in this way, let us look first at the profile function of the
energy density. If we take the following iteration formula:
\begin{equation}\label{crr}
\varrho(T_n(\tau))=\varrho(\tau)  D_n^2(\tau) \ ,
\end{equation}
together with Eqs.~(\ref{pert1},\ref{pert2}), then it is
straightforward to derive the following asymptotic formulas for
large arguments ($n\gg 1$):
\begin{equation}\label{edas1}
\varrho(nT+\tau_++\varepsilon D_n^{-1}(\tau_+)) \cong
\varrho(\tau_++\varepsilon)D_n^2(\tau_++\varepsilon) \ ,
\end{equation}
\begin{equation}\label{edas2}
\varrho(nT+\tau_-+\varepsilon) \cong \varrho(\tau_-+\varepsilon
D_n(\tau_-))D_n^2(\tau_-+\varepsilon D_n(\tau_-)) \ .
\end{equation}
The above formulas explain the formation of travelling narrow
packets in the energy density $T_{00}(t,x)$
\cite{law,lambrecht,dodonov} . The profile function
$\varrho(\tau)$ is concentrated around spots of the positive
periodic trajectory. From Eqs.~(\ref{edas1},\ref{edas2}) one can
estimate easily the width and the height of peaks. The height
grows like $D_n^2(\tau_+)$, and the width diminishes like
$D_n^{-1}(\tau_+)$. The values of the energy density far off the
peaks decreases according to $D_n^2(\tau_-)$.

The total energy can be analyzed using the following formula:
\begin{equation}\label{cte}
E(T^\star_n(\tau_0))=\int_{T_{n-1}(\tau_0)}^{T_n(\tau_0)} \ d\tau
\ \varrho(\tau) = \int_{f(\tau_0)}^{\tau_0} \ d\tau \
\varrho(\tau) D_n(\tau) \ .
\end{equation}
We can calculate the evolution of the total accumulated energy
using only initial data and the cumulative D\"{o}ppler factor. It
is apparent that an initial shape of the classical field inside
the cavity is of minor importance.  Let us test our method with an
example of sinusoidal cavity wall motion:
\begin{equation}\label{sin}
    L(t)=L+\Delta L \sin{(\omega t)} \ .
\end{equation}
Obviously, we are to assume that $\Delta L<L$ and $\omega\Delta
L<1$. The parametric resonance frequencies are $\omega_N=N\pi/L$,
where $N$ is the order of the resonance. It is easy to find
periodic particle trajectories. First, we read all return points
of the mirror from Eq.~(\ref{pc}): $\sin(\omega_N \tau^\star)=0$.
They correspond to the following starting points of positive and
negative particle trajectories:
\begin{equation}
\tau_{+m}=\frac{(-N+2m+1)L}{N} \ \ \ , \ \ \
\tau_{-m}=\frac{(-N+2m)L}{N} \ ,
\end{equation}
where $m=0,1,...,N-1$. The corresponding values of the cumulative
D\"{o}ppler factors can be calculated:
\begin{equation}
D_n(\tau_{\pm m})=\left(\frac{1\pm\omega_N\Delta
L}{1\mp\omega_N\Delta L}\right)^n
 \cong \exp{(\pm 2n\omega_N\Delta
L)}  \ \quad {\rm for} \ \ \Delta L\ll L \ .
\end{equation}
There are $N$ travelling peaks in the energy density. All the
peaks have the same height and the same width. The corresponding
time evolution of peaks agrees with the results of quantized
version \cite{dodonov}. The formation of peaks is basically a
classical phenomenon.

The resonance instability in the field theory appears not only for
finely tuned frequencies. We obtain usually some band structure.
Our method enable us to get insight into off resonant behavior of
the cavity system as well. Therefore, we assume some perturbation
of the resonant frequency: $\omega=\omega_N+\Delta \omega$. The
equation for return points is now:
\begin{equation}
\sin(\omega \tau^\star)=-\frac{L\Delta\omega}{\omega\Delta L} \ .
\end{equation}
The solutions for return points  exist provided that:
\begin{equation}\label{bs}
    \frac{\Delta \omega}{\omega}<\frac{\Delta L}{L} \ .
\end{equation}
This condition has been obtained numerically for  quantized
version in \cite{wegrzyn2}. It defines the band structure of the
parametric resonance \cite{arnold}. For small perturbations, the
structure defined by Eq.~(\ref{bs}) is typical and resembles those
obtained in classical mechanics. The cumulative D\"{o}ppler factor
at starting points yields:
\begin{equation}
D_n(\tau_{\pm m})=\left(\frac{1\pm\sqrt{(\omega\Delta
L)^2-(L\Delta\omega)^2}}{1\mp\sqrt{(\omega\Delta
L)^2-(L\Delta\omega)^2}}\right)^n .
\end{equation}
For off resonant vibrations of the cavity, the formation of narrow
peaks and the energy growth take a longer time. Let us conclude
our classical analysis. Within the framework of our method the
calculations are straightforward for any type of cavity motion.
The parametric resonance is related to the existence of periodic
trajectories in a two-dimensional cavity billiard. As it was
already noticed in \cite{cole}, there is no need for periodicity
of cavity motion. We need only periodic returns of the cavity to
the unperturbed position. For typical cavity motions, the growth
of the total energy is exponential with time. The cumulative
D\"{o}ppler factor describes all details of the formation of
narrow wave packets and the energy growth. Essentially, many
details of the resonant enhancement of classical electromagnetic
radiation inside a vibrating cavity match the results of quantum
theory.

Finally, we discuss vibrating optical cavities in the quantum
field theory. Apart from the interference effects which we have
already demonstrated in the classical theory of the cavity, we are
to account for instability of the quantum vacuum and possibility
for particle production. The vacuum expectation value of the
energy density is given by \cite{moore}:
\begin{equation}
\langle T_{00}(t,x)\rangle = \varrho(t+x)+\varrho(t-x) \ ,
\end{equation}
where
\begin{equation}
\varrho(\tau)=-\frac{\pi}{48}
\dot{R}^2(\tau)-\frac{1}{24\pi}S[R](\tau) \ .
\end{equation}
The second term is responsible for particle production. It is
defined by the Schwartz derivative:
\begin{equation}
S[R]=\frac{\stackrel{...}{R}}{\dot{R}}-\frac{3}{2}
\left(\frac{\stackrel{..}{R}}{\dot{R}}\right)^2
\ .
\end{equation}
The crucial information is contained in the phase function, which
obeys the following Moore's equation (being a counterpart of
kinematical laws Eqs.(1,2,\ref{dbc})):
\begin{equation}\label{me}
    R(\tau)-R(f(\tau))=2 \ .
\end{equation}
We see that our billiard function is useful for the quantum case
as well. It is straightforward to derive the following recursive
relation:
\begin{equation}\label{rr}
\varrho(T_n(\tau))=\varrho(\tau)D^2_n(\tau)+A_n(\tau)D^2_n(\tau) \
,
\end{equation}
where the cumulative D\"{o}ppler factor is given again by
Eq.~(\ref{cdf}) and the cumulative conformal anomaly contribution
can be calculated from the following formulae:
\begin{equation}\label{cac}
    A_n(\tau)= \frac{1}{24\pi} S[T_n](\tau)=
     - \frac{1}{24\pi}\sum_{k=1}^n
    D_k^{-2}(\tau)S[f](T_k(\tau)) \ .
\end{equation}
The total energy can be calculated now from:
\begin{equation}\label{qte}
E(T^\star_n(\tau_0))=\int_{T_{n-1}(\tau_0)}^{T_n(\tau_0)} \ d\tau
\ \varrho(\tau)
 = \int_{f(\tau_0)}^{\tau_0} \ d\tau \ \left[
\varrho(\tau)+ A_n(\tau) \right] D_n(\tau) \ .
\end{equation}
In comparison with the classical formula Eq.~(\ref{crr}), aside
from the D\"{o}ppler factor there appears a new additive term on
the right hand side. It represents the conformal anomaly of the
theory.  Most of classical results, like the band structure around
resonance frequencies, the formation and the shape of travelling
packets in the energy density, the exponential growth of the total
accumulated energy, can be reproduced in the quantum case as well.
These features are to be calculated from the cumulative
D\"{o}ppler factor. The  novelty of the quantum description is the
conformal anomaly.  We show that for the lowest resonance channel,
the anomalous mechanism of energy growth clashes with the
resonance enhancement of the initial vacuum fluctuations.

Let us refer again to the sinusoidal motion of the cavity
(\ref{sin}). We consider motions with resonance frequencies
$\omega_N$.  At the beginning, the cavity is static and empty. It
corresponds to the following condition:
\begin{equation}
\varrho(\tau)=-\frac{\pi}{48L^2}  = -\frac{\omega_1^2}{48\pi} \ \
\quad {\rm for} \ \ \tau \in [-L,L] \ .
\end{equation}
The above value is just the static Casimir energy density which is
present even the cavity is at rest. We can compute the cumulated
anomaly contribution at starting points of positive trajectories
directly from Eq.~(\ref{cac}):
\begin{equation}
    A_n(\tau_{+m})=\frac{\omega_N^2}{48\pi} \frac{1}{1-\omega_N^2
    \Delta L^2} \left[ 1 - D_n^{-2}(\tau_{+m})  \right] \ .
\end{equation}
We see that for  a long time limit (here large $n$) and small
amplitudes, the initial density contribution $\varrho(\tau)$ and
the anomaly contribution $A_n(\tau)$ cancel each other out for
$N=1$. It explains why the lowest resonance (called sometimes
${\it semi-resonance}$ \cite{dodonov}) is suppressed. No particles
are produced inside a cavity. For higher resonance frequencies
$N>1$, the anomalous mechanism surpasses resonant enhancement of
the negative energy of the initial state. We obtain again the
exponential growth of the total energy and the formation of
travelling narrow wave packets.

The first exact analytical solution corresponding to a vibrating
cavity system was presented in \cite{law} for the resonance
channel $N=2$ and generalized in \cite{wu} for higher resonances.
The solutions correspond to a family of mirror trajectories given
by:
\begin{equation}\label{lawwu}
    L(t)=L+\frac{1}{\omega_N} \left\{ \arcsin{\left[
    \sin{\frac{\omega_N \Delta L}{2}} \cos{(\omega_N t)} \right] -
\frac{\omega_N \Delta L}{2} } \right\}
 \ . \end{equation}
The picture of resonantly enhanced radiation is similar there
except of the fact that the energy happens to grow quadratically
with time. Such a power-like growth of the total energy is
observed at the boundary of the frequency band Eq.~(\ref{bs}). For
trajectories Eq.~(\ref{lawwu}), the billiard function can be
derived exactly:
\begin{equation}
    f(\tau)=\frac{2}{\omega_N} {\rm arccot} \left[
    \cot{\frac{\omega_ (\tau-L)}{2}} - 2 \tan{\frac{\omega_N
    \Delta L}{2}} \right] - L \ ,
\end{equation}
where the branch of the multivalued function arccot should be
properly chosen to have the billiard function increasing. It is
straightforward to calculate the cumulative D\"{o}ppler factor and
the cumulative anomaly contribution:
\begin{equation}
    D_n(\tau)=1+2n^2\tan^2{\frac{\omega_N \Delta L}{2}} \left[ 1 -
    \cos{\left( \omega_N (\tau+L) \right)} \right]
     +2n\tan{\frac{\omega_N \Delta
    L}{2}} \sin{\left( \omega_N (\tau+L) \right)} \ ,
\end{equation}
\begin{equation}
    A_n(\tau)=\frac{\omega_N^2}{48\pi} \left[ 1- D_n^{-2}(\tau)
    \right] \ .
\end{equation}
There are $N$ periodic particle trajectories corresponding to
starting points $\tau_{0m}=(-N+ 2m+1) L/N$ where $m=0,1,...,N-1$.
But all $D_n(\tau_{0m})=1$. The periodic trajectories are neither
positive nor negative. There is no D\"{o}ppler factor since the
reflections occur at the turning points where the mirror stops. We
obtain the corresponding energy density:
\begin{equation}
\varrho(\tau)=- \frac{N^2 \pi}{48\L^2} + \frac{(N^2-1)
\pi}{48\L^2} \left\{ 1+
 2n^2\tan^2{\frac{\omega_N \Delta L}{2}}
\left[ 1 - (-1)^N \cos{\left( \omega_N \tau \right)} \right]
 -2n(-1)^N  \tan{\frac{\omega_N \Delta L}{2}}
 \sin{\left( \omega_N \tau \right)} \right\}^{-2}
\end{equation}
In this case, it is easy to show that the total energy grows
quadratically with time, and the heights and widths of travelling
wave packets are proportional to $\tau^4$ and $\tau^{-2}$
respectively.

\section{Summary}

In summary, we have studied the resonance behavior of the
electromagnetic field inside a vibrating one-dimensional (linear)
cavity. Our approach exploits fully the analogy with the classical
billiard of massless particles. The parametric resonance is
determined by the existence of periodic particle trajectories. The
billiard function can be used to establish recursive relations for
all physical quantities. The resonant exponential enhancement and
the concentration of energy into narrow wave packets is described
 in both classical and quantum field theories of cavities
 referring to the same aggregated
 D\"{o}ppler factor. It is enough to calculate this factor, even
 numerically, in some finite interval of the length related to the
 period of cavity oscillations. A purely quantum phenomenon of
 particle production from the vacuum is described by the anomalous
 contribution in recursive relations. For higher resonances, this
 contribution dominates and the evolution of the system is
 insensitive to the initial state of the quantum field inside a
 cavity. However, for higher resonances the band structure defined
 by  Eq.~(\ref{bs}) is squeezed. From experimental point of view, the
 inequality Eq.~(\ref{bs}) challenges someone to secure either fine
 tuning  or big enough amplitude with gigahertz frequencies.
 The resonance  amplification and photon production from thermal
 field fluctuations \cite{lambrecht2}, being practically more
 feasible, can be studied in the same way as well.

The opto-mechanical resonance in  vibrating cavities (dynamical
Casimir effect) is the subject of numerous studies. The analysis
of real, three-dimensional models shows that each photon mode can
be truly described by some one-dimensional model. We have
presented a new and simple method to analyze resonance solutions
in such fundamental models. Our new results: the mechanism of
particle production and the mechanism of resonance amplification
of radiation are separated and clarified, exact formulas for
off-resonant solutions are given, the band structure is described,
the puzzle with the lack of the amplification for the lowest
resonant frequency is solved, it is also explained when the
instability of the system (the energy growth) is either
exponential or power-like. Moreover, the correspondence with
classical optics and classical mechanics is presented here from a
new point of view as far as the quantum parametric resonance is
concerned.



\begin{thebibliography}{99}

\bibitem{berges} J. Berges, J. Serreau, Phys. Rev. Lett. {\bf 91},
111601 (2003).
\bibitem{hirooka} H. Hiro-oka, H. Minakata, Phys. Rev. {\bf C 64},
044902 (2001).
\bibitem{lambrecht} A.Lambrecht, M.T.Jaeckel, S.Reynaud,
Phys. Rev. Lett. {\bf 77}, 615 (1996).
\bibitem{moore} G.T. Moore. J. Math. Phys. {\bf 11}, 2679, (1970).
\bibitem{fulling} S.A. Fulling, C.W. Davies, Proc. R. Soc. Lond. A {\bf
348}, 393 (1976).
\bibitem{dodonov} V.V.Dodonov, Modern Nonlinear Optics, Part 1,
p.309-394
 (Advances in Chemical Physics, Volume 119, ed. by M.W.Evans, Wiley, N.Y., 2001);
  quant-ph/0106081.
\bibitem{bordag} M.Bordag, U.Mohideen, V.M.Mostepanenko, Physics Reports {\bf
353}, 1-205 (2001).
\bibitem{meplan} O. M\'{e}plan, C. Gignoux, Phys. Rev. Lett. {\bf 76}, 408 (1996).
\bibitem{nagatani} Y.Nagatani, K.Shigetomi, Phys. Rev. {\bf A 62},
022117 (2000).
\bibitem{law} C.K.Law, Phys. Rev. Lett. {\bf 73}, 1931 (1994).
\bibitem{dittrich} J.Dittrich, P.Duclos, P.\u{S}eba, Phys. Rev.
{\bf E 49}, 3535 (1994).
\bibitem{wegrzyn} P. W\c{e}grzyn, T. R\'og, Acta Phys. Polon. {\bf
32}, 129 (2001).
\bibitem{cole} C.K. Cole, W.C. Schieve, Phys.Rev. {\bf A 53}, 4495
(1995).
\bibitem{wegrzyn2} P. W\c{e}grzyn, T. R\'og, Acta Phys. Polon. {\bf
34}, 3887 (2003).
\bibitem{arnold} V. I. Arnold, {\it Mathematical Methods Of
Classical Mechanics} (Springer, New York, 1978), Sec. 25 and Sec.
42.
\bibitem{wu} Y.Wu, K.W.Chan, M.C.Chu, P.T.Leung, Phys.Rev. {\bf A 59},
1662 (1999).
\bibitem{lambrecht2} A.Lambrecht, M.T.Jaeckel, S.Reynaud,
Europhys. Lett. {\bf 43}, 147 (1998).






\end{thebibliography}
\end{document}